# Toward an African Agenda for AI Safety

Samuel T. Segun[1], Rachel Adams[1], Ana Florido[1], Scott Timcke[2], Jonathan Shock[3], Leah Junck[1], Fola Adeleke[1], Nicolas Grossman[1], Ayantola Alayande[1], Jerry John Kponyo[4], Matthew Smith[5], Dickson Marfo Fosu[4], Prince Dawson Tetteh[4], Juliet Arthur[4], Stephanie Kasaon[6], Odilile Ayodele[7], Laetitia Badolo[8], Paul Plantinga[7], Michael Gastrow[7], Sumaya Nur Adan[9], Joanna Wiaterek[10], Cecil Abungu[11], Kojo Apeagyei[12], Luise Eder[13], Tegawende Bissyande[14]

[1]Global Center on AI Governance, [2]Research ICT Africa, [3]University of Cape Town, [4]Kwame Nkrumah University of Science and Technology (KNUST), [5]International Development Research Centre, [6]AI Action Lab, [7]Human Sciences Research Council (HSRC), [8]Niyel, [9]Oxford Martin AI Governance Initiative, [10]Equiano Institute, [11]University of Cambridge, [12]Qhala, [13]University of Oxford, [14]University of Luxembourg

## Abstract

This paper, authored by researchers and practitioners from Africa, maps the continent's distinctive risk profile, from deep-fake–fuelled electoral interference and data-colonial dependency to compute scarcity, labour disruption and disproportionate exposure to climate-driven environmental costs. Artificial intelligence (AI) technologies are set to transform African economies, polities and ecosystems. While major benefits are promised to accrue, the availability, development and adoption of AI also mean that African people and countries face particular AI safety risks, from large-scale labour market disruptions to the nefarious use of AI to manipulate public opinion. To date, African perspectives have not been meaningfully integrated into global debates and processes regarding AI safety, leaving African stakeholders with limited influence over the emerging global AI-safety governance agenda. Analysing current national strategies and the new AU Continental AI Strategy, we identify severe gaps in technical capacity, governance and participation: only 26.8% of states measured showed any concrete activity on safety, accuracy or reliability. While there are Computer Incident Response Teams on the continent, none hosts a dedicated AI Safety Institute or office. We propose a five-point action plan centred on (i) a policy approach that foregrounds the protection of the human rights of those most vulnerable to experiencing the harmful socio-economic effects of AI; (ii) the establishment of an African AI Safety Institute; (iii) promote public AI literacy and awareness; (iv) development of early warning system with inclusive benchmark suites for 25+ African languages; and (v) an annual AU-level AI Safety & Security Forum.

**Introduction**

Safety concerns have been at the forefront of recent global conversations regarding the capabilities and effects of frontier AI[1] development. These concerns, and the debates that have ensued, have been accelerated by the rapid emergence of new AI systems, including generative and agentic AI models, which raise new and heightened concerns about the large-scale risks and harms their use or misuse could lead to.[2,3] To address these concerns, in 2023, the first AI Safety Summit was held at Bletchley Park in the UK, bringing together political and tech leaders from around the world to determine how best to safeguard the use of frontier AI systems, and catapulting the field of "AI Safety" to a policy issue of global concern. The Summit led to the establishment of AI Safety Institutes in the US, UK, EU, Japan, Singapore, South Korea and Canada, and further AI Safety Summits held in South Korea (2024), Kenya (2024) and France (2025). A number of these institutes have subsequently been renamed AI Security Institutes.[4,5] Although there have been calls to establish a regional AI safety taskforce under the African AI Council[6] and an AI safety institute in Africa,[7] no dedicated research and policy centre on AI safety has been established on the Continent yet. Policy interest in AI Safety institutes has surged precisely as major technology companies scale back internal safety measures, prioritizing deployment of increasingly advanced AI models to compete for market share.[8]

Against the aforementioned backdrop, it is important to define what AI Safety really is. "AI safety" is a scientific field concerned with identifying and addressing the security, ethical, socioeconomic, environmental, technical, and existential risks and harms of frontier AI models.[9] To date, AI safety research has sought to understand the technical capabilities as well as weaknesses of artificial general intelligence (AGI), the risks they pose to society, and the technical and policy solutions needed to mitigate these risks. The field is thought to contribute to ensuring that advanced AI systems are designed and deployed in ways that prevent unintended harms and align with human values.[10]

---

[1] "Frontier AI" refers to the development of AI models and systems that break new ground in technological possibility, and are accordingly at the forefront of the techno-scientific field of AI.
[2] Naudé, W., Dimitri, N. The race for an artificial general intelligence: implications for public policy. *AI & Soc* **35**, 367–379 (2020). https://doi.org/10.1007/s00146-019-00887-x.
[3] Gyevnar, B., & Kasirzadeh, A. (2025). AI safety for everyone. *Nature Machine Intelligence*, 1–12.
[4] AI safety refers to preventing any unintended harm from AI systems, from technical failures to large-scale socio-economic impacts. AI security is narrower: it protects AI models, data and infrastructure against malicious interference (e.g., adversarial attacks, model theft) and also covers the use of AI for broader cyber-defence.
AI alignment is a technical sub-field that makes an AI system's objectives and behaviour reliably match human intentions and values, averting goal-mis-specification or deceptive optimisation as capability scales. See: https://cloudsecurityalliance.org/blog/2024/03/19/ai-safety-vs-ai-security-navigating-the-commonality-and-differences.
[5] Claudia Wilson, (2025 February 14), AI Safety Is Becoming AI Security. Center for AI Policy. https://www.centeraipolicy.org/work/ai-safety-is-becoming-ai-security
[6] Joanna Wiaterek, Cecil Abungu, and Chinasa T. Okolo, (2025 May 7). Building regional capacity for AI safety and security in Africa. *Brookings Institute*. https://www.brookings.edu/articles/building-regional-capacity-for-ai-safety-and-security-in-africa/
[7] Rumman Chowdhury, 'What the Global AI Governance Conversation Misses' *Foreign Policy* (September, 2024). https://foreignpolicy.com/2024/09/19/ai-governance-safety-global-majority-internet-access-regulation/.
[8] Frank Ryan, Niki Iliadis, and George Gor, "Weaving a Safety Net: Key Considerations for How the AI Safety Institute Network Can Advance Multilateral Collaboration" (The Future Society, 2024).

AI systems pose risks and harms are relevant globally, and are also shared by African countries. Some AI risks apply disproportionately to less developed countries, or to the Global South. However, in many ways the African context is distinct from other marginalised regions. One clear distinction is Africa's common history, which includes the era of precolonial cultures, the colonial era, postcolonial state-building, and currently being positioned between competing geopolitical powers, including powerful actors in the AI landscape. Against a long history in which the sovereignty of African peoples has been systematically undermined by global powers, whether colonial states or neo-colonial tech giants, Africa has every historical and policy reason to focus on sovereignty[11] in both its technological trajectories and its international relations.

Our focus on the African continent takes into account that Africa is a highly diverse region linguistically, culturally, economically, and environmentally. The dynamics of AI policy are evidently diverse too. However, the continent also has a significant common history and geopolitical position. This history has cultivated African solidarity, which has been expressed through the creation of the African Union and other Pan-African organisations. With respect to AI policy, the AU therefore plays an important role in co-ordinating voices, interests, and policy approaches across the African continent, in order to have a greater impact at the global level.

At present, there is limited contribution of African perspectives to the discourse on AI Safety. Given that the risks that frontier AI models pose are universal (although their distribution and impact will not be equal[12]), and that the science to understand, identify and mitigate these risks is still emerging, it is paramount that deliberations on AI safety reflect globally diverse expertise and evidence. African leaders and stakeholders can provide important perspectives on AI safety that can advance rich global discussions and secure Africa's participation in ensuring the safe use of AI technology on the continent.[13]

---

[9] Baldridge, D., Coleman, B., and Demanuele, A. (2024 May 29). The terminology of AI regulation: Ensuring "safety" and building "trust". *Schwartz Reisman Institute for Technology and Society*. https://srinstitute.utoronto.ca/news/terminology-regulation-safety-trust
[10] There has been some debate on this. See https://michaelnotebook.com/xriskbrief/index.html.
[11] Digital/technological sovereignty in this context refers to Africa's ability to control its own technology infrastructure and digital ecosystem, ensuring independence and self-reliance.
[12] Rachel Adams, *The New Empire of AI: The Future of Global Inequality*, (Polity, 2024).
[13] Anthony, A., Munga, J., and Appaya, S. (2024 November, 28). From the margins to the center: Africa's role in shaping AI governance. *World Bank Blog*.

The omission of African perspectives in these debates can lead to governance measures that do not address the complex safety and security challenges posed by AI which the continent faces.[14] Similarly, with major influential states and actors increasingly focused on the technical dimensions of AI safety, there is a risk that the socio-economic impacts of AI may be overlooked in global governance efforts, which could be detrimental to regions like Africa that are more vulnerable to these types of risks. Given the priority being placed by governments and development partners on the use of AI to fast-track socio-economic development in Africa, it is critical that the safety considerations of these technologies are considered and addressed alongside their adoption.

In this white paper, we adopt an understanding of AI safety as a socio-technical discourse focused on the significant threats AI systems pose to the security of both humanity and the environment. The paper sets out key considerations for AI safety from an African perspective and advances a five-step action plan for building safe and secure AI in the region. Section 1 examines the major risks to human and environmental security that AI poses within an African context. This section follows the categorization of risks adopted in the International AI Safety Report 2025, which is currently considered the gold standard in the field. Section 2 maps the current state of AI governance relating to AI safety in Africa, and Section 3 sets out the five-point action plan for an African research and policy agenda on AI safety, including highlighting opportunities to integrate African perspectives into global debates and decision-making. African perspectives are vital for ensuring that collective global governance of the large-scale risks of AI is comprehensive and robust, drawing from diverse insights to craft inclusive and agile governance mechanisms which protect all lives, communities and environments from the harms of AI.

**Section 1: AI Risks in Africa**

While AI adoption may be less widespread across Africa, the risks it poses to African communities are no less urgent, and potentially more severe given existing vulnerabilities and resource constraints[15]. The African continent is home to a complex environment, both in terms of the risks to the safety and security of persons on account of various factors, as well as the technical integrity of digital systems. As AI enters this landscape, there is

---

https://blogs.worldbank.org/en/governance/from-the-margins-to-the-center--africa-s-role-in-shaping-ai-gove.

[14] This is not unique to the African context. The Global Index on Responsible AI found that the majority of countries around the world lack the mechanisms to protect human rights at risk from AI. www.global-index.ai.

[15] Cecil Abungu, Marie Victoire Iradukunda, Duncan Cass-Beggs, Aquila Hassan, Raqda Sayidali. (2024 December 16). Why Global South Countries Need to Care About Highly Capable AI. Centre for International Governance Innovation. Paper No. 311.
https://www.cigionline.org/publications/why-global-south-countries-need-to-care-about-highly-capable-ai/

considerable work to be done to anticipate, identify and address the challenges that have and may emerge.

The AI safety discourse can be classified into two – immediate and remote. Immediate operational risks are short-term risks, while remote risks are long-term, potentially catastrophic risks. The 2025 International AI Safety Report[16] discuss these short-term and long-term risks putting into three broad categories of AI risks namely:

1. Risks from malicious use: These are risks associated with the intentional misuse or weaponization of AI systems to harmful purposes.
2. Risks from malfunction: These are risks caused by the technical shortcomings of AI systems which can have physical, psychological, reputational, financial, and legal consequences.
3. Systemic risks: These are risks that stem from the broader societal impact of AI use and not just the use of large scale or frontier models.

The following section discusses these risks as they pertain to the African continent. It highlights considerations that African leaders and stakeholders face regarding AI safety.

### 1. Risks from Malicious Use

- Risks to democracy and human rights

With AI-driven technologies like facial recognition, the risk of misuse for surveillance purposes is high, given that many African leaders do not have a track record of tolerating opposition. For example, activists in Zimbabwe fear that the use of facial recognition cameras deployed in Bulawayo could potentially be used for surveillance and hamper digital rights[17]. Similarly, activists in Uganda, noting that the government has not had a track record of upholding human right values, have raised concerns about the police use of facial recognition cameras, which the police claimed has helped them track down and arrest more than 836 suspects[18]. Additionally, AI tools and software owned or run by foreign firms could enable authoritarian regimes to suppress dissent surveillance. An example of this was the Nigerian government's request to Twitter in 2022 seeking backend access to enable it to delete user posts that were considered inappropriate or anti-government[19].

---

[16] Yoshua Bengio et al. (2025 January). The International Scientific Report on the Safety of Advanced AI. AI Action Summit.
https://assets.publishing.service.gov.uk/media/679a0c48a77d250007d313ee/International_AI_Safety_Report_2025_accessible_f.pdf

[17] Farai Shawn Matiashe. (2024 February 14). Zimbabwe: Digital rights activists fear misuse of surveillance cameras in Bulawayo. The Africa Report.
https://www.theafricareport.com/336723/zimbabwe-digital-rights-activists-fear-misuse-of-surveillance-cameras-in-bulawayo/

[18] Stephen Kafeero. (2020 November 27). Uganda is using Huawei's facial recognition tech to crack down on dissent after anti-government protests. *Quartz*.
https://qz.com/africa/1938976/uganda-uses-chinas-huawei-facial-recognition-to-snare-protesters

Additionally, AI-generated content has been used in the recent past to subvert democratic processes, delegitimize electoral institutions, and influence outcomes[20]. For example, AI bots have been used to spread disinformation and in other cases used to generate or shape public narratives at scale. Countries in Africa are particularly prone to such subversion[21, 22] given the growing influence of social media platforms as a medium of information dissemination. With fragile democratic institutions and the continued rise in mobile internet penetration across Africa, with one of its primary uses being access to social media, AI generated content raises serious concerns for the continent. For instance, during Kenya's 2022 elections, AI-altered audio clips and social media bots spread false narratives, exacerbating ethnic tensions[23].

Open AI's 2025 report on its work to disrupt malicious uses of its models documented the attempt to use ChatGPT generated "short comments and long-form articles" to influence electoral outcomes in the just concluded Ghanaian presidential election[24]. The company banned ChatGPT accounts of a self-acclaimed youth organization with active social media accounts which used ChatGPT to generate comments on its posts.

- Risks of manipulation of public opinions

The manipulation of public opinion through social media and other channels has always been a challenge for digital governance. However, with AI, this problem has been amplified. AI-generated deepfakes and disinformation not only threaten Africa's democratic processes but also lead to manipulation of public opinions. This could have a disastrous health and safety impact, making governance problematic. For example, during the COVID-19 Pandemic, disinformation about the transmission, treatment protocol and conspiracy theories made managing the pandemic difficult for many countries in Africa[25]. This manipulation of public opinions leads to people refusing treatment or even attacking healthcare workers as was the case in the August 2018 and June 2020 Ebola outbreak in the Democratic Republic of Congo[26].

---

[19] People's Gazette .(2022 January 12). Twitter agreed to register in Nigeria, give us backend access to delete citizens' posts: Buhari Regime.
https://gazettengr.com/twitter-agreed-to-register-in-nigeria-give-us-backend-access-to-delete-citizens-posts-buhari-regime/

[20] Samson Itodo. (2024 March 26). Artificial Intelligence and the integrity of African elections. International Institute for Democracy and Electoral Assistance.
https://www.idea.int/news/artificial-intelligence-and-integrity-african-elections

[21] Tina Gerhäusser & Martina Schwikowski. (2025 February 23). AI disinformation could threaten Africa's elections. Digital World.
https://www.dw.com/en/ai-disinformation-could-threaten-africas-elections/a-71698840

[22] Scott Timcke, (2025 May 21). Focus on the democratic fundamentals: What comes before the promise of AI in African elections. Research ICT Africa.
https://researchictafrica.net/research/focus-on-the-democratic-fundamentals-what-comes-before-the-promise-of-ai-in-african-elections/

[23] Peter Mwai (2022 July 10). Kenya elections 2022: The misinformation circulating over academic qualifications. BBC News. https://www.bbc.com/news/62070665

[24] Ben Nimmo, Albert Zhang, Matthew Richard, Nathaniel Hartley. (2025 February). Disrupting malicious uses of our models: an update. Open AI.
https://cdn.openai.com/threat-intelligence-reports/disrupting-malicious-uses-of-our-models-february-2025-update.pdf

Similarly, the manipulation of public opinion can have a far-reaching effect on the social cohesion of many of Africa's consolidating democracies, causing strife, ethnic tensions and straining weak alliances. For example, in Ethiopia, deepfakes and AI-generated hate speech on social media worsened ethnic conflicts during the Tigray War[27] hampering the peace-making process and has continued to worsen social relations in the country[28]. These risks continually increase because the access to government and its institutions remains distant to many[29], especially those in rural settlements and because AI has lowered the barriers to creating and disseminating these types of disinformation.

- Risks of external influence

Today, Africa's engagement with AI is often mediated by external actors, and in this case American and Chinese companies. Both China and the US, through proxies, have strategies and activities that have the potential to undermine African AI sovereignty, in what amounts to something of a digital land grab. These strategies reflect the different cultures and politics of the respective geopolitical powers.

China, for example, is emerging as a dominant influence in digital infrastructure, AI models, and the energy needed to power them. Chinese cloud dominance raises sovereignty and risk management challenges for African regulators. Chinese tech giants such as Alibaba and Huawei are expanding their presence by offering cloud services and investing heavily in data centers across the continent. Huawei's plans to invest $430 million in data centers in Africa, and Alibaba's cloud services in South Africa illustrate the scale of this influence. Similarly, US companies like Microsoft in partnership with G42 are becoming increasingly prominent players in the field of (sovereign) cloud infrastructure and AI development, which

---

poses a risk of creating dependencies.[30] In 2024, Microsoft announced a $1 billion comprehensive digital ecosystem initiative for Kenya.[31] Such expansion raises several questions about whose interest such systems serve and whether they are adequately attuned to the continent's unique socio-economic and cultural contexts[32].

A significant risk of relying on external entities for AI development is the challenge of technological dependency, which consequently can undermine Africa's sovereignty. For example, a report in 2020 accused China of eavesdropping on sensitive conversations in the African Union headquarters[33] in Addis Ababa, Ethiopia, and sharing such information with Chinese companies operating in Africa[34]. The espionage was alleged to have been carried out through the cameras provided by Huawei. Although the Chinese government has denied the allegations, this poses a concern as such information sharing could be justified under China's surveillance laws. Chinese corporations operating in Africa such as TikTok, DeepSeek and RedNote are bound to comply with China's surveillance laws, which could mandate the sharing of users' information with the Chinese government[35]. The point is not to single out or label any country as a malicious actor, but rather to highlight the AI safety considerations that African policy-makers must understand. .

- Risks of militarization, abuse and loss of control

There are growing concerns about the militarization of AI technologies and the effect it would have on Africa. The African Union (AU) has flagged the lack of regulatory frameworks for lethal autonomous weapons[36] as a concern. Although national AI strategies and policy guides are currently being developed across Africa, very few address the potential use of AI in the defence and security industry. Kenya has however hosted the Africa regional consultation of the Responsible use of Artificial Intelligence in Military (REAIM)[37], showing its commitment as one of the 60 countries that have endorsed its non-binding "call to action" document.[38]

---

[30] Microsoft Corporation. (2024). AI in Africa: Meeting the Opportunity.
https://msblogs.thesourcemediaassets.com/sites/5/2024/01/AI-in-Africa-Meeting-the-Opportunity.pdf

[31] Microsoft Corporation. (2024 May 22). Microsoft and G42 announce $1 billion comprehensive digital ecosystem initiative for Kenya.
https://news.microsoft.com/source/2024/05/22/microsoft-and-g42-announce-1-billion-comprehensive-digital-ecosystem-initiative-for-kenya/

[32] Salami (2024). Artificial intelligence, digital colonialism, and the implications for Africa's future development. Data & Policy. 6. https://doi.org/10.1017/dap.2024.75

[33] Raphael Satter. (2020 December 16). Exclusive-Suspected Chinese hackers stole camera footage from African Union - memo. Reuters.
https://www.reuters.com/article/world/exclusive-suspected-chinese-hackers-stole-camera-footage-from-african-union-me-idUSKBN28Q1EA/

[34] Dipanjan Roy Chaudhury. (2020 December 27). China uses Huawei cameras to spy on African Union headquarters. The Economic Times.
https://economictimes.indiatimes.com/news/international/world-news/china-uses-huawei-cameras-to-spy-on-african-union-headquarters/articleshow/79981099.cms?utm_source=contentofinterest&utm_medium=text&utm_campaign=cppst

[35] Aaron Katersky, Kaitlyn Morris, Kate Holland, and Alexandra Myers .(2025 February 5). DeepSeek coding has the capability to transfer users' data directly to the Chinese government. ABC News.
https://abcnews.go.com/US/deepseek-coding-capability-transfer-users-data-directly-chinese/story?id=118465451

[36] The African Union. (2024 April 17). Remarks by Dr. Alhaji Sarjoh Bah, Director of Conflict Management Directorate in the Political Affairs, Peace and Security Department, to the Workshop on Autonomous Weapons System: An ECOWAS Perspective.
https://au.int/en/speeches/20240417-remarks-dr-alhaji-sarjoh-bah-director-conflict-management-directorate-political

Autonomous drones, used in counter-terrorism efforts raise concerns about accountability and accidental civilian harm[39]. Similarly, cases of attacks on military bases and camps in Nigeria and Mali[40] by malicious non-state actors like violent extremist organizations and terrorists raise legitimate concern about the hijack of AI-enabled military technologies[41]. With troubling reports of arms trafficking in the Sahel region[42], the need for regulating the development and use of AI-enabled military technologies is paramount.

There are additional risks posed by private military contractors like Wagner Group operating in Africa, which employs AI systems without oversight; this could lead to escalating existing conflict, as has been seen in the Sahel region[43].

Lastly, there is growing concern about terrorist groups like Jama'at Nasr al-Islam wal Muslimin (JNIM), the Islamic State (IS), Boko Haram Al-Shabaab, who operate on the continent, using publicly available open-source AI tools for propaganda and radicalization. For example, a report by the Tech Against Terrorism's Open-Source Intelligence (OSINT) identified and archived over "5,000 pieces of AI-generated content produced by terrorist and violent extremist actors"[44]. Additionally, these groups could leverage open-source AI

---

[37] The Kenyan Ministry of Defence. (2024 June 11). Responsible AI in the Military Domain. https://www.mod.go.ke/news/responsible-ai-in-the-military-domain/

[38] Brandon Vigliarolo. (2023 February 17). The International military AI summit ends with a 60-state pledge. *The Register*. https://www.theregister.com/2023/02/17/military_ai_summit/

[39] Crisis Group. (2023 December 20). Türkiye's Growing Drone Exports. https://www.crisisgroup.org/europe-central-asia/western-europemediterranean/turkiye/turkiyes-growing-drone-exports

[40] Wedaeli Chibelushi & Paul Njie. (2024 September 17). Al-Qaeda-linked group says it was behind Mali attack. BBC News. https://www.bbc.com/news/articles/ce8d996x1r0o

[41] Sahara Reporters. (2025 February 15). Boko Haram Terrorists Attack Nigerian Army Base, Kill Three Soldiers, Steal Gun Trucks In Borno. https://saharareporters.com/2025/02/15/breaking-boko-haram-terrorists-attack-nigerian-army-base-kill-three-soldiers-steal-gun

[42] Genevieve Jess. (2025 July 12). Arms Trafficking: Fueling Conflict in the Sahel. The International Affairs Review. https://www.iar-gwu.org/print-archive/ikjtfxf3nmqgd0np1ht10mvkfron6n-bykaf-ey3hc-rfbxp-dpte8-klmp4

[43] Center for Preventive Action. (2024 October 23). Violent Extremism in the Sahel. Global Conflict Tracker. https://www.cfr.org/global-conflict-tracker/conflict/violent-extremism-sahel

[44] Tech Against Terrorism. (2023, November). Early terrorist experimentation with generative artificial intelligence services. [Report]. https://techagainstterrorism.org/hubfs/Tech%20Against%20Terrorism%20Briefing%20-%20Early%20terrorist%20experimentation%20with%20generative%20artificial%20intelligence%20services.pdf

models to enhance and protect their identities and privacy, frustrating counter-terrorism efforts[45].

- Risk of cybercrime and financial malpractice

Although many African states have data protection laws, their enforcement remains an issue. This creates room for data and financial exploitation, like in the case of predatory lending in Africa[46]. AI-powered phishing and voice cloning target Africa's booming fintech sector. For example, in Ghana, "sakawa" cybercriminals use deepfakes to impersonate relatives in mobile money scams[47]. Kenya recently recorded about 840 million cyber threats between October and December of 2024[48]. These attacks as well as AI-driven financial crimes have been on the increase in Africa[49].

Risk of cybercrimes and exploitation are also exacerbated by advances in the capabilities of AI-assisted programming tools. These tools can find vulnerabilities and hack into government websites that already have weak cybersecurity features. Similarly, AI-assisted programming tools used to generate codes may inherit vulnerabilities from their training data, often without full context or security awareness. This creates a security blind spot where AI-generated code might contain exploitable flaws that developers do not fully understand or scrutinize and can unintentionally introduce vulnerabilities.

- Risks to the safety and security of women and girls

AI tools contribute to a rise in online gender-based violence, including the creation and distribution of abusive content such as deepfakes credibly manipulating images, voices and actions. This not only leads to individual harms, but it colours public perceptions and creates more openings for further exploitation of those who are marginalised across different intersections, such as race and ethnicity, gender, beliefs, ability, literacy, socioeconomic status and geographic location. For example, accessible Generative AI tools are creating new vectors for gender-based harm, from non-consensual intimate imagery to algorithmic amplification of misogyny and systematic exploitation of women's online presence.[50] It is worth noting that these harms are extensions of long-standing gender inequalities.

---

[45] Samuel Segun. (2025). The Global Security Risks of Open-Source AI Models. United Nations Office for Disarmament Affairs Blog. https://disarmament.unoda.org/responsible-innovation-ai/blog/#_ftn15

[46] Stephen Mutie. "Debt, extraction and predation." *African political economy in the twenty-first century* (2023): 341-365.

[47] Yushawu, A., & Jaishankar, K. (2025). Sakawa in Ghana: The Influence of Weak Ties on Economic Cybercrime Offender Networks. *Deviant Behavior*, 1–21. https://doi.org/10.1080/01639625.2025.2459681

[48] Elijah Ntongai. (2025 February). Kenya records 840 million cyber threats in 3 months, says criminals used AI to increase attacks. MSN. https://www.msn.com/en-xl/africa/kenya/kenya-records-840-million-cyber-threats-in-3-months-says-criminals-used-ai-to-increase-attacks/ar-AA1xTBS4?ocid=finance-verthp-feeds&apiversion=v2&noservercache=1&domshim=1&renderwebcomponents=1&wcseo=1&batchservertelemetry=1&noservertelemetry=1

[49] Mohammed Yusuf. (2025 February 13). AI-driven biometric fraud surges in Africa, fueling financial crimes. Voice of America. https://www.voanews.com/a/ai-driven-biometric-fraud-surges-in-africa-fueling-financial-crimes/7973925.html

During elections, women in journalism and politics are specifically targeted and utilised to tarnish reputations, spread falsehoods and intimidate[51] AI-generated abusive content has also intensified online gender-based violence more broadly, impacting their wellbeing and the ways in which girls and women feel free to participate in online spaces.[52] Similarly, unauthorised surveillance has been weaponised to invade women's privacy and publicly shame them, rendering patriarchal control an invisible systemic threat.[53]

Risks of gendered AI misuse are currently not being sufficiently addressed through regulation with the Global Index on Responsible AI (link) highlighting significant gaps across the globe in that regard.[54] Yet, there are notable regional women-led innovations that manage to utilise the abilities of AI to respond to the everyday realities of GBV violence, such as HerSafeSpace in Nigeria, Zuzi in South Africa, and Ame in Botswana, supporting survivors and promoting prevention.

## 2. Risks from Malfunction

- Reliability & safety

AI reliability remains a standout ethical issue. AI innovators in Africa are developing novel solutions to address country-specific needs across multiple sectors such as education, healthcare, financial services, and agriculture. Although this constitutes progress, the reliance on models primarily trained on Euro-American datasets raise questions about the reliability and accuracy of these solutions[55].

Africa's infrastructure limitations exacerbate the challenges of reliability in AI. For instance, the inadequacy of quality and structured data means that scaling AI solutions could 12present challenges in reliability[56]. Furthermore, infrastructure deficits like unreliable electricity could create frequent system downtimes during critical diagnostic processes for AI diagnostic or triaging tools[57], which could raise questions about their reliability and use for medical predictions.

---

[50] Scott Timcke and Zara Schroeder (2024, August) Engagement at What Cost? Examining the intersection of social media, Generative AI and gender-based violence, Research ICT Africa, https://researchictafrica.net/research/engagement-at-what-cost-examining-the-intersection-of-social-media-generative-ai-and-gender-based-violence/
[51] https://eprints.soton.ac.uk/499751/
[52] https://researchictafrica.net/research/engagement-at-what-cost-examining-the-intersection-of-social-media-generative-ai-and-gender-based-violence/
[53] https://wougnet.org/download/gendered-aspects-of-and-its-impact-on-data-protection-for-women-and-girls-in-africa/
[54] https://genderequalitybrief6.tiiny.site/
[55] Karen Haoarchive. (2018 December 2). AI has a culturally biased world view that Google has a plan to change. MIT Technology Review. https://www.technologyreview.com/2018/12/02/138843/ai-has-a-culturally-biased-worldview-that-google-has-a-plan-to-change/

AI use in education has been a potent use case globally, and the challenge of reliability in light of model hallucination is not uncommon. AI systems are known to sometimes misrepresent facts or historical evidence[58]. AI use in education, as personalised tutors, have raised concerns for teachers who question its reliability[59]. LLMs, on which many AI education tools are built, exhibit unreliability in mathematics, raising concerns about their application and use[60]. African innovators face an added challenge because many countries lack strong research infrastructure and local data. Without this, scientific and AI models may fail to reflect local needs, leading to misaligned or ineffective solutions. Additionally, it seems unlikely that the current architecture of large language models will be able to solve these challenges.[61]

- Bias and fairness

The absence of diverse and representative data could create algorithmic bias in AI systems[62]. As previously mentioned, current frontier AI models are trained on Western data[63], creating a challenge when used out of such contexts. Context biases have been experienced by AI innovators and practitioners in Africa, requiring fine tuning and other methods used to mitigate such biases[64].

---

[56] Gary Drenik. (2023 August 15). Data Quality For Good AI Outcomes. Forbes. https://www.forbes.com/sites/garydrenik/2023/08/15/data-quality-for-good-ai-outcomes/

[57] Guo J, Li B. The Application of Medical Artificial Intelligence Technology in Rural Areas of Developing Countries. Health Equity. 2018 Aug 1;2(1):174-181. doi: 10.1089/heq.2018.0037

[58] Nicola Jones. (2025 January 21). AI hallucinations can't be stopped — but these techniques can limit their damage. Nature. https://www.nature.com/articles/d41586-025-00068-5

[59] FE News Editor. (2024 March 21). Two thirds of teachers think AI is too unreliable for the classroom, new report finds. https://www.fenews.co.uk/education/two-thirds-of-teachers-think-ai-is-too-unreliable-for-the-classroom-new-report/

[60] Satpute, Ankit, Noah Gießing, André Greiner-Petter, Moritz Schubotz, Olaf Teschke, Akiko Aizawa, and Bela Gipp. "Can llms master math? investigating large language models on math stack exchange." In *Proceedings of the 47th international ACM SIGIR conference on research and development in information retrieval*, pp. 2316-2320. 2024.

[61] Anthropic. (2025 June 20) Agentic Misalignment: How LLMs could be insider threats. https://www.anthropic.com/research/agentic-misalignment

[62] Merlyna Lim and Ghadah Alrasheed. (2021 May 16). Beyond a technical bug: Biased algorithms and moderation are censoring activists on social media. Carleton Newsroom. https://newsroom.carleton.ca/story/biased-algorithms-moderation-censoring-activists/

[63] Vered Shwartz. (2024 February 13). Artificial intelligence needs to be trained on culturally diverse datasets to avoid bias. The Conversation. https://theconversation.com/artificial-intelligence-needs-to-be-trained-on-culturally-diverse-datasets-to-avoid-bias-222811

[64] James Manyika, Jake Silberg and Brittany Presten. (2019 October 25). What Do We Do About the Biases in AI? Harvard Business Review. https://hbr.org/2019/10/what-do-we-do-about-the-biases-in-ai

Algorithmic biases in facial recognition have been one of the clearest examples documented in the recent past. For example, Joy Boulamwini and Timnit Gebru's research on gender bias in facial recognition software revealed an error rate of 0.8% for light-skinned men and 34.7% for dark-skinned women[65]. The results not only reveal a bias in terms of race but one of gender as well. Although these were not systems built in Africa, it underscores the concern about biases amplified by AI systems.

Large language models (LLMs), as documented in various studies, often display biases related to race, gender, religion, and political viewpoints.[66] Another study showed that ChatGPT predominantly mirrors American values,[67] a trend that could be problematic when used in other cultural contexts.[68] Historical biases found in data also tend to be exacerbated by AI systems. For example, research on maternal healthcare in Southern Africa shows that maternal health AI solutions are at risk of "replicating biases due to datasets that do not reflect culturally relevant practices and knowledge"[69]. In an anecdotal account of the impact of automated decision-making software in loan financing, historical data excluding black South African women amplified loan application rejections when race or gender is accounted for[70].

Content moderation algorithms present another example that has shown the effects of AI biases in Africa. For example, in 2020, Meta's content moderation algorithm in the Middle East and North Africa incorrectly flagged posts 77% of the time, raising serious concerns about its reliability in detecting hate speech and extremist content[71]. In some cases, particularly during periods of conflict, Meta's content moderation failed to flag hate speech and violent content. Notable examples include Ethiopia during the Tigray War and the Hawassa Riots.[72]

---

[65] Buolamwini, Joy, and Timnit Gebru. (2018). "Gender shades: Intersectional accuracy disparities in commercial gender classification." In *Conference on fairness, accountability and transparency*, pp. 77-91.
[66] Motoki, F., Pinho Neto, V. & Rodrigues, V. More human than human: measuring ChatGPT political bias. Public Choice 198, 3–23 (2024). https://doi.org/10.1007/s11127-023-01097-2
[67] Yihan Cao, Siyu Li, Yixin Liu, Zhiling Yan, Yutong Dai, Philip S. Yu, Lichao Sun (2023). A Comprehensive Survey of AI-Generated Content (AIGC): A History of Generative AI from GAN to ChatGPT. Arxiv Preprint. arXiv:2303.04226
[68] Scott Timcke and Hanani Hloman. (2024). Decoding the Ballot: How Might AI Reshape Democracy on the African Continent? Research ICT Africa. https://researchictafrica.net/research/how-might-ai-reshape-democracy-on-the-african-continent/
[69] Min'enhle Ncube. ( 2024 August). Incomplete Chronicles: Unveiling Data Bias In Maternal Health. Mozilla. https://assets.mofoprod.net/network/documents/Incomplete_Chronicles._AIMZ_report.revised.pdf
[70] Naadiya Moosajee. ( 2019 March 14). Fix AI's racist, sexist bias. Mail & Guardian. https://mg.co.za/article/2019-03-14-fix-ais-racist-sexist-bias/
[71] Jillian C. York, Paige Collings, and David Greene. (2025 January 9). Meta's New Content Policy Will Harm Vulnerable Users. If It Really Valued Free Speech, It Would Make These Changes. Electronic Frontier Foundation. https://www.eff.org/deeplinks/2025/01/metas-new-content-policy-will-harm-vulnerable-users-if-it-really-valued-free#:~:text=Part%20of%20the%20problem%20is,77%20percent%20of%20the%20time.
[72] Prithvi Iyer. (2025 April 21). What a New Study Reveals About Content Moderation in Tigray. *Tech Policy Press*. https://www.techpolicy.press/what-a-new-study-reveals-about-content-moderation-in-tigray/

## 3. Systemic Risks

- Labour risks

AI poses a unique economic risk for Africa. AI-guided automation, for example, may disrupt proven developmental pathways.[73] About 40% of the global economy is exposed to a potential AI disruption, according to data from the International Monetary Fund (IMF)[74], with advanced economies better positioned to buffer the risks that AI poses to the job market. With developments in agentic models[75] to manage tasks like customer relations, supply chain management, data analysis etc., there is an increasing likelihood that job loss will become more widespread. The 2025 Future of Jobs Report by the World Economic Forum raises more concerns for Sub-Saharan Africa, indicating a significant exposure to AI disruption. The report finds that, in the coming years, approximately two-thirds of new workforce entrants will be supplied by India and Sub-Saharan African countries.[76] Currently, entry-level jobs have been identified as most susceptible to advances in AI[77], and this vulnerability is likely to be particularly pronounced in many African countries.

Many English-speaking African countries have strategically positioned themselves over the last decade to be the destination for business process outsourcing. For example, Kenya is home to almost two million digital workers according to the Business Process Outsourcing Association of Kenya (BPOAK), and in Ghana, BPO contributes about $200m to its economy.[78] These countries had an advantage from an abundance of labour, which when combined with infrastructure for commodity exports, has attracted manufacturing and service industries.[79] However, this advantage is being eroded as AI-guided automation makes it more cost effective for firms in wealthier countries to 'de-couple' their operations from the

---

[73] Acemoglu, D. (2021). Harms of AI (Working Paper No. 29247). National Bureau of Economic Research. https://doi.org/10.3386/w29247

[74] Mauro Cazzaniga, Florence Jaumotte, Longji Li, Giovanni Melina, Augustus J Panton, Carlo Pizzinelli, Emma J Rockall, and Marina Mendes Tavares. "Gen-AI: Artificial Intelligence and the Future of Work", *International Monetary Fund*. 2024, 001 (2024). https://doi.org/10.5089/9798400262548.006

[75] Agentic AI models are AI systems that are autonomously capable of pursuing, planning, and executing complex tasks independent of human intervention with access to specific productivity tools..

[76] Attilio Di Battista, Sam Grayling, Ximena Játiva, Till Leopold, Ricky Li, Shuvasish Sharma, Saadia Zahidi (2020 January). The Future of Jobs Report. *World Economic Forum*. Geneva: Switzerland. https://reports.weforum.org/docs/WEF_Future_of_Jobs_Report_2025.pdf

[77] Till Leopold, (2025 April 30). How AI is reshaping the career ladder, and other trends in jobs and skills on Labour Day. World Economic Forum. https://www.weforum.org/stories/2025/04/ai-jobs-international-workers-day/#:~:text=AI%20stands%20as%20one%20of,facing%20the%20labour%20market%20today.&text=While%20170%20million%20new%20jobs.collar%2C%20entry%2Dlevel%20roles

[78] Ben Payton. (2025 May 23). Africa aims to crack the $300bn business process outsourcing market. African Business.https://african.business/2025/05/technology-information/africa-aims-to-crack-300bn-business-process-outsourcing-market

[79] Morris, M., Fessehaie, J. (2024) The industrialisation challenge for Africa: Towards a commodities based industrialisation path. Journal of African Trade 1: 25–36. https://doi.org/10.1016/j.joat.2014.10.001

global economy.[80] This shift disproportionately affects African workers whose roles are susceptible to automation, particularly those with limited specialized training. For example, the African telecom giant MTN introduced its Zigi chatbot and virtual assistant which manages customer queries and has had an impact on the number of call center representatives[81].

Women, particularly, are at an increased risk of being made redundant by automation in Africa.[82] Similarly, gig workers who recently embraced the platform economy are especially vulnerable to these developments.[83] Beyond the labour market, African countries may lose opportunities for industrial growth and economic development. This situation is further complicated by unequal access to AI technologies themselves, creating a challenge where African countries must simultaneously manage the disruption of their economic sectors while being encouraged to build the skills and infrastructure for a future with AI.

- Environmental risks

Environmental risks posed by AI systems are a cause for concern. Carbon dioxide ($CO_2$) emissions are projected to increase as more frontier models are developed.[84] More troubling is that this number would increase as the development and use of these models increase. For example, the International Monetary Fund projects that additional $CO_2$ emissions associated with AI could reach about 1.7 gigatons from 2025 – 2030.[85] However, it should be noted that the training and inference emissions differ as do their energy requirements, and newer models appear to seek ways to judiciously use energy. [86]

Researchers from the University of Massachusetts, Amherst, performed a life cycle assessment for training various LLMs. The results indicates that the process could cause the emission of up to 626,000 pounds of $CO_2$[87], which is roughly equivalent to the annual greenhouse emissions of about 6 average US households or 62 average cars. The environmental cost is also evident when the models are fine-tuned[88] and grow proportionally to the model size. The impact of AI-induced $CO_2$ is noteworthy because Africa faces a disproportionate burden from climate change[89].

---

[80] Rana Foroohar, R. (2023, August 7). The Truth about Decoupling, Financial Times, https://www.ft.com/content/2843370b-3494-464e-83e6-263679873aa3; Timcke, S. (2024). AI Regulation Paradigms and the Struggle over Control Rights. The Bulletin of Technology & Public Life. https://doi.org/10.21428/bfcb0bff.ae40312a.
[81] Charles Dominic & Peng Boris Akebuno. (2025 February 20). What Jobs Will ChatGPT and AI Replace in the Near Future? Tech Culture Africa. https://techcultureafrica.com/what-jobs-will-ai-replace
[82] https://genesis.imgix.net/uploads/files/Preparing-for-AI-in-the-BPO-and-ITES-Sector-in-Africa.pdf
[83] Tichafara Dinika, A. (2024) Digital Labour Mirage: Dissecting Utopias in Africa's Gig Economy, PhD Dissertation, Universität Bremen, https://media.suub.uni-bremen.de/handle/elib/8082.
[84] Bender, E. M., Gebru, T., McMillan-Major, A., & Shmitchell, S. (2021, March). On the dangers of stochastic parrots: Can language models be too big?🦜. In *Proceedings of the 2021 ACM conference on fairness, accountability, and transparency* (pp. 610-623).
[85] Christian Bogmans, Patricia Gomez-Gonzalez, Ganchimeg Ganpurev, Giovanni Melina, Andrea Pescatori, and Sneha D Thube. "Power Hungry: How AI Will Drive Energy Demand", *IMF Working Papers* 2025, 081 (2025), accessed May 8, 2025, https://doi.org/10.5089/9798229007207.001
[86] Andy Masley. (2025 January 13). Using ChatGPT is not bad for the environment. The Weird Turn Pro. https://andymasley.substack.com/p/individual-ai-use-is-not-bad-for
[87] Strubell, Emma, Ananya Ganesh, and Andrew McCallum. (2020). "Energy and policy considerations for modern deep learning research." In Proceedings of the AAAI conference on artificial intelligence, vol. 34, no. 09, pp. 13693-13696.

Equally important are concerns around energy and water use. AI systems consume a significant amount of electricity and the data centers which provide the computing access they use also require fresh water for cooling. Data from the International Energy Agency (IEA) indicates that data centers consumed nearly 460 terawatt-hours (TWh) of electricity globally, with the number expected to increase substantially with the development of more complex models and larger data centers.[90]

Water use is also a challenge for countries where fresh water remains a scarce resource. A report from the Lawrence Berkeley National Laboratory estimated that U.S. data centers directly consumed approximately 21.2 billion liters of water in 2014, increasing to 66 billion liters in 2023.[91] This number is expected to increase in light of the recent developments in AI. At the moment, nearly 230 million Africans are confronting water scarcity, and up to 460 million will inhabit water-stressed regions by the end of 2025, according to data from Africa Relief[92]. Countries on the continent such as South Africa, Kenya, Ethiopia, Somalia, still face water insecurity, which could see pushbacks from communities requesting that they be excluded as locations for large data centers. However, this might change in the future due to advancements in closed-loop cooling systems.[93] Also, drought-prone African countries may face challenges concerning water usage and agricultural activities due to the water demands of data centers.

With rapid advances made in AI, there has been an increased demand for specialised hardware like graphics processing units (GPUs), tensor processing units (TPUs), hard drives,

---

[88] Karen Hao. (2019 June 6). Training a single AI model can emit as much carbon as five cars in their lifetimes. MIT Technology Review.
https://www.technologyreview.com/2019/06/06/239031/training-a-single-ai-model-can-emit-as-much-carbon-as-five-cars-in-their-lifetimes/

[89] World Meteorological Organization. (2024 September 02). Africa faces disproportionate burden from climate change and adaptation costs.
https://wmo.int/news/media-centre/africa-faces-disproportionate-burden-from-climate-change-and-adaptation-costs#:~:text=%E2%80%9CAfrica%20faces%20disproportionate%20burdens%20and,water%20resources%2C%20and%20overall%20socio%2D

[90] *International Energy Agency*. Electricity 2024: Analysis and forecast to 2026.
https://iea.blob.core.windows.net/assets/6b2fd954-2017-408e-bf08-952fdd62118a/Electricity2024-Analysisandforecastto2026.pdf (2024).

[91] Shehabi, Arman, Alex Hubbard, Alex Newkirk, Nuoa Lei, Md Abu Bakkar Siddik, Billie Holecek, Jonathan Koomey, Eric Masanet, and Dale Sartor. "2024 United States Data Center Energy Usage Report." (2024).

[92] Africa Relief (2024 June). Water Scarcity In Africa: Causes & Solutions.
https://africarelief.org/blogs/africa-water-crisis

[93] Naduvilakath-Mohammed, F. M., R. Jenkins, G. Byrne, and A. J. Robinson. "Closed loop liquid cooling of high-powered CPUs: A case study on cooling performance and energy optimization." *Case Studies in Thermal Engineering* 50 (2023): 103472.

IoT and edge devices and sensors. As older hardware devices are phased out for newer generation, they contribute to the electronic waste (e-waste) production. Data from the Global E-Waste Monitor published by the UN Institute for Training and Research shows that a record 62 million tonnes (Mt) of e-waste was produced in 2022, which is up 82% from 2010 and on track to rise another 32%, to 82 million tonnes in 2030[94].

As demand for AI increases, the need for newer and more capable hardware devices to replace older and slower generation become necessary. This raises another potential AI-induced environmental risk for Africa. For example, it is estimated that the Agbogbloshie landfill in Accra, Ghana, receives an estimated 15,000 tons of e-waste[95].

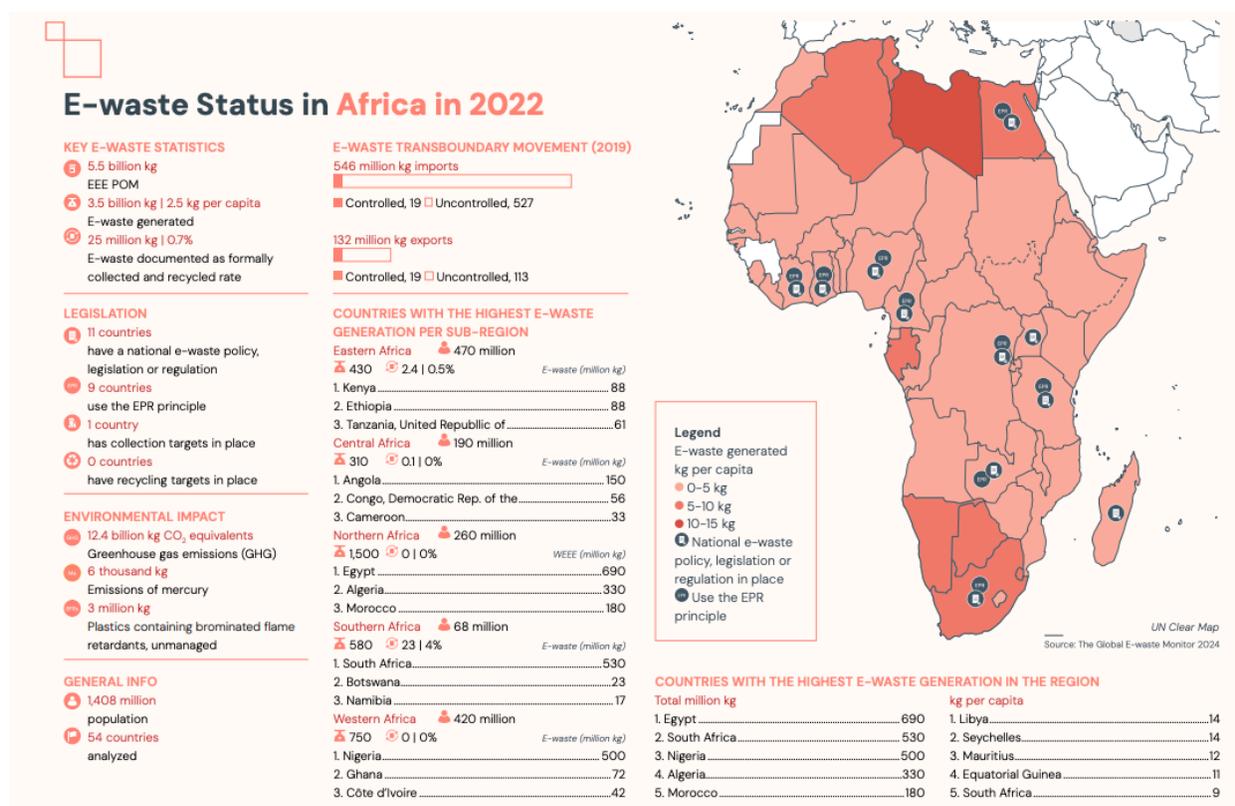

Source: Global e-Waste Monitor 2024[96]

---

[94] Baldé, C.P., Kuehr, R., Yamamoto, T., McDonald, R., D'Angelo, E., Althaf, S., Bel, G., Deubzer, O., Fernandez-Cubillo, E., Forti, V. and Gray, V., 2024. The global e-waste monitor. *United Nations University (UNU), International Telecommunication Union (ITU) & International Solid Waste Association (ISWA), Geneva/Bonn*, pp.1-147.https://api.globalewaste.org/publications/file/297/Global-E-waste-Monitor-2024.pdf

[95] Jonathan Lambert. (2024 October 5). Stunning photos of a vast e-waste dumping ground — and those who make a living off it. NPR. https://www.npr.org/sections/goats-and-soda/2024/10/05/g-s1-6411/electronics-public-health-waste-ghana-phones-computers

[96] Same as above

E-waste poses health risks for millions of Africans who work in the informal recycling sector, hence requiring tough legislation to regulate its operations. Currently, only 11 African countries have a national e-waste legislation[97], policy or regulation with Cameroon being the first[98]. Only Egypt, Ghana, Nigeria, Côte d'Ivoire, Cameroon, Tanzania, Rwanda and Uganda, South Africa, Zambia, Malawi, have clear national policies on e-waste, with Kenya developing a draft legislation.

## Section 2: AI Safety in Africa – State of the Art in African AI Governance Efforts

The notion of AI safety cannot be separated from broader ethical and contextual considerations, including the expressions of many treaties, charters and soft laws that articulate the aspiration to transform Africa into a globally competitive, integrated, and prosperous continent with shared prosperity, unity, and good governance.[99] Deploying AI systems that are not informed by the continent's cultural, linguistic and societal nuances is a big risk. For instance, Raji and Sholademi (2024) addresses the issue of biases noting that predictive policing algorithms, developed in vastly different socio-legal contexts, can exacerbate racial and socio-economic profiling if deployed without significant adaptation.[100] Similarly, Salami (2024) cautions that AI-driven credit scoring systems might entrench financial exclusion if they fail to account for the informal economic practices that characterize much of Africa's economy. These examples show why we need to ensure AI systems in Africa are not only technically safe but also ethically aligned with the values and needs of African communities[101]; if rolled out, they must be trained in such a way that they do not further entrench biases that may have propagated from the Global North.[102] Even if these systems are built on contextually relevant data, their deployment is not advisable due to the inherent difficulty for AI to adequately address such social prediction problems.

---

[97] Same as above
[98] Africa News. (n.d.). Cameroon: The NGO recycling and processing e-waste.
https://www.africanews.com/2022/11/01/cameroon-the-ngo-recycling-and-processing-e-waste/
[99] African Union. (2015). Agenda 2063: The Africa we want. African Union Commission.
https://au.int/sites/default/files/documents/33126-doc-framework_document_book.pdf African Union. (2007). African Charter on Democracy, Elections and Governance. African Union.
https://au.int/sites/default/files/treaties/36384-treaty-african_charter_on_democracy_elections_and_governance.pdf African Union. (2001). The New Partnership for Africa's Development (NEPAD). African Union.
https://www.nepad.org/publication/new-partnership-africas-development-nepad
[100] Raji, I., & Sholademi, D. B. (2024). *Predictive Policing: The Role of AI in Crime Prevention*. International Journal of Computer Applications Technology and Research, 13(10), 66–78.
[101] Segun, Samuel. "Are certain African ethical values at risk from artificial intelligence?." *Data & Policy* 6 (2024): p68.
[102] Models that are fine-tuned locally with local data stand a better chance of being both locally accurate and relevant, as well as less likely to cause bias. See this policy brief for a further discussion of how open source LLMs can be used by local innovators to train models on local data, and the challenges associated therein:
https://www.globalcenter.ai/research/implications-of-open-source-large-language-models-for-responsible-ai-development-in-africa.

419The cost of computing resources presents a critical barrier to Africa's AI development and safety. The continent's compute power and infrastructure is negligible compared to the rest of the world.[103, 104] For example, according to the AI Hub for Sustainable Development, about 75% of the world's supercomputers are hosted in countries in the Global North with less than 1% hosted in Africa; the continent also accounts for 2% of the global data centers[105], one of the lowest compared to other regions, although there have been new investments to address this gap. This disparity means that even if African countries were to mobilize all their current computational resources toward the development of larger AI models, they would lag behind competitors in North America, Europe, and East Asia. This 'computational poverty' effectively locks Africa out of sovereign AI development, forcing dependence on foreign AI technologies and perpetuating a new form of digital colonialism. Crucially for this white paper, the limitations of computing resources make it difficult for Africans groups to conduct complex automated evaluations of the capabilities of frontier AI models, which constitutes a core part of the work of many AI safety institutes.[106]

The integration of AI technologies in Africa presents a challenge that, if not managed well, could echo historical patterns of colonialism. Consider that while AI offers opportunities for language preservation and cultural documentation, there is also the potential risk of intersecting forms of layered marginalization. The current push by some tech companies to incorporate African languages into AI models raise questions about control, selection, and linguistic diversity. If some African languages receive the lion's share of attention due to commercial viability, there is a risk that new linguistic hierarchies will be created. The risk extends beyond language to encompass deeper cultural paradigms. The challenge lies in ensuring that AI development in Africa becomes a tool for cultural empowerment rather than vectors for hierarchy or cultural erosion.

While some African stakeholders have been represented within the broader network of the AI safety institutes,[107] AI safety is still an emerging area of governance in Africa. Although some African countries have focused on developing national AI strategies and policies, few address the primary concerns about AI safety. The Global Index on Responsible AI, which surveyed 19 areas of responsible AI in 138 countries, shows that Africa has the lowest average score in the area 'Safety, Accuracy and Reliability'.[108] Out of 41 African countries assessed, only 11 provided evidence of activities to advance AI safety, accuracy and reliability in at least one of the three pillars of evidence (Government frameworks, Government actions, and Non-state actors). Within 30 African countries, no evidence could be found in relation to local efforts to advance discussions on AI safety. Below, we unpack the results of this index, highlighting the gaps in Africa's AI readiness.

---

[103] Mutiso, R. M. (2024, March 28). AI in Africa: Basics Over Buzz. Science Vol 383, Issue 6690, https://www.science.org/doi/10.1126/science.ado8276

[104] Lehdonvirta, V., Wú, B., & Hawkins, Z. (2024, October). Compute North vs. Compute South: the uneven possibilities of compute-based AI governance around the globe. In *Proceedings of the AAAI/ACM Conference on AI, Ethics, and Society* (Vol. 7, No. 1, pp. 828–838).

[105] AI Hub for Sustainable Development. (n.d). Building the Africa Green Compute Coalition, https://www.aihubfordevelopment.org/green-compute-coalition

[106] See also the efforts of the Africa Green Compute Coalition: https://www.aihubfordevelopment.org/green-compute-coalition.

[107] Kenya has participated in the International Network of AI Safety Institutes represented by its Special Envoy for Technology, Philip Thigo. See: Mabonga, P. (2024 November 24). Ambassador Thigo Leads Kenya's Participation in Historic AI Safety Network Launch in the US. *Kenya News Agency*. https://www.kenyanews.go.ke/ambassador-thigo-leads-kenyas-participation-in-historic-ai-safety-network-launch-in-the-us/

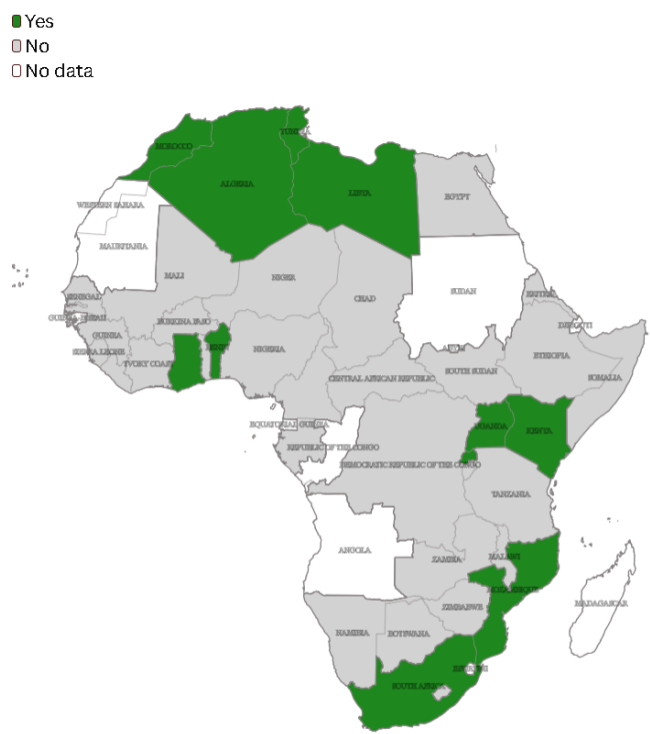

Source: Global Index on Responsible AI

At the time of writing, Kenya is the only African country to demonstrate evidence of activities across all three pillars (Government frameworks, Government actions, and Non-state actors) in relation to AI Safety, Accuracy and Reliability. Within the "Emerging Technologies" section of the [Kenya Digital Masterplan](), the need for a future National AI Strategic Plan that ensures the safety and security of AI systems is highlighted. It stresses that before AI systems are broadly deployed, there must be assurance they operate safely, securely, and in a controlled, well-defined, and well-understood manner—calling for additional research to overcome the challenges of making AI systems reliable, dependable, and trustworthy. This commitment is embedded within the broader strategic initiative, "Develop Strategy for Adoption of Smart Technologies," which outlines the government's plan to develop a robust AI framework as part of Kenya's overall digital transformation agenda.

---

[108] Adams, R., Adeleke, F., Florido, A., de Magalhães Santos, L. G., Grossman, N., Junck, L., & Stone, K. (2024). Global Index on Responsible AI 2024 (1st Edition). South Africa: Global Center on AI Governance. [www.global-index.ai](www.global-index.ai).

Other notable countries engaged in efforts to address AI safety and security include Ghana, Morocco, and Rwanda. In Ghana, the government is working on developing a policy framework to govern the use of AI which highlights Ghana's plan to ensure that AI is safe, reliable and accurately used in the country.[109]

As part of the Moroccan government's efforts, the National Human Rights Council (CNDH) launching the Rabat Declaration published a report focused on Artificial Intelligence and Digital Citizenship, which emphasized AI safety.[110] The document promotes AI safety by advocating for a "Human Rights by Design" approach, ensuring AI systems protect citizens from harm. A scientific committee was established to implement the recommendations, reinforcing efforts to integrate human rights safeguards into AI governance.

The Rwanda National AI Policy emphasizes the need for ethical and safety precautions to ensure AI solutions benefit citizens and do not cause harm.[111] It promotes ethical AI development, a robust data strategy and infrastructure. It suggests improving AI-ready data, strengthening data governance, and establishing procurement guidelines to support trustworthy AI adoption.

In addition, the South African government has also made important strides in setting out the need for robust measures to ensure the technical integrity and safety of AI. In 2024, the Department of Digital Technologies and Communications published the National AI Policy Framework for public comment. The Framework includes a provisional section on Safety and Security calling for the implementation of "robust cybersecurity protocols to safeguard AI systems" and the development of "frameworks to identify and mitigate risks associated with AI".[112]

---

[109] Source: Connielove Mawutornyo Dzodzegbe. (2023 June 8). Government working on a framework to regulate use of AI – Communications Minister. Joy Online. https://www.myjoyonline.com/government-working-on-a-framework-to-regulate-use-of-ai-communications-minister/

[110] National Human Rights Council (CNDH). (n.d.). Technology, Artificial Intelligence & Human Rights. https://www.cndh.ma/en/technology-artificial-intelligence-human-rights

[111] Republic of Rwanda Ministry of ICT and Innovation. (2022). The National AI Policy. https://www.minict.gov.rw/index.php?eID=dumpFile&t=f&f=67550&token=6195a53203e197efa47592f40ff4aaf24579640e

[112] These developments were not included in the 1st Edition of the Global Index on Responsible AI as they fell outside of the data collection period, which ran from November 2021 to November 2023. The SA National AI Policy Framework was published for public comment in 2024.

**The African Union's Continental AI Strategy & AI Safety**

The African Union has shown a commitment to AI safety within the recently published AU Continental AI Strategy.[113] This is a welcome effort to define a distinctive African perspective on AI, one that prioritizes inclusion and equity, as well as the safe use of AI technologies. Comparably, the recent Declaration at the Global AI Summit on Africa mentioned little with regard to AI safety.[114] The Continental Strategy acknowledges the dual nature of AI, that is, its ability to drive social progress and its potential to worsen existing inequalities or introduce new risks. It highlights the need for clear safeguards and oversight mechanisms to address societal, ethical, security, and legal challenges associated with AI, setting out 15 areas of policy action. The 11th refers to the "adoption and implementation of technical standards to ensure the safety and security of AI systems across the Continent". Under this provision, the Continental AI Strategy calls for:

- African governments to "develop innovative and agile regulatory instruments and frameworks to address safety and security challenges of advanced and complex AI Systems";
- The hosting of an "Annual Conference on AI Safety and Security in Africa"; and
- The establishment of an "Expert Group to assess AI's impact on peace and security on the continent". The AU has subsequently appointed a set of experts to constitute an Advisory Group on AI and its impact on peace, security, and governance.[115]

While the strategy is a big step forward, its success is still contingent on the willingness and ability of AU Member States to translate its principles into actionable policies and enforceable regulations. The strategy also offers limited emphasis on technical safety measures, such as robust detection systems for misinformation or public interest algorithms even though it acknowledges risks like disinformation.[116]

Despite progress at the continental level, there are considerable lags amongst countries in Africa in terms of developing AI regulatory legislations. While some have put together national AI policies, frameworks or guidelines, others still lag behind. However, it's noteworthy that many African countries have put considerable effort into establishing foundational laws that can guarantee AI safety such as data protection, cybersecurity, consumer protection, intellectual property and competition laws[117]. 36 of the continent's 54 countries have all enacted data protection laws[118], and 39 of them have dedicated cybersecurity/cybercrime laws.[119] These foundational instruments could be used as starting points to better understand, strengthen, and develop key aspects of AI Safety.

---

[113] African Union. (2024). *Continental Artificial Intelligence Strategy*. Addis Ababa: African Union Commission. Retrieved from https://au.int/en/documents/20240809/continental-artificial-intelligence-strategy
[114] https://c4ir.rw/docs/Africa-Declaration-on-Artificial-Intelligence.pdf.
[115] Ibid.
[116] Chege, G. (2024 September). The AU Continental AI Strategy: Concrete Safety Proposals or High-tech Hype? *Ilina Program*. https://ilinaprogram.org/2024/09/27/the-au-continental-ai-strategy-concrete-safety-proposals-or-high-tech-hype/.

Attention to the needs and priorities of the African continent with regard to AI safety governance remain significantly under-addressed in the broader conversations especially by governments and international alliances. This is because much of the research and policy work on AI safety is being carried out by diaspora experts and non-state actors – civil society and academia.

**Section 3: Recommendations**

In this section we outline a five-pillar action plan for AI safety in Africa which includes (i) a policy approach foregrounding the protection of the human rights of those most vulnerable to experiencing the harmful socio-economic effects of AI; (ii) the establishment of an African AI Safety Institute dedicated to studying the risks and impact of AI on the continent; (iii) a proposal to promote public AI literacy and awareness across Africa; (iv) the development of early warning system with inclusive benchmark suites for 25+ African languages; and (v) an annual AU-level AI Safety & Security Forum. These five pillars are targeted at policymakers (regional, governments), private sector and non-profit (for funding and investments), and educational institutions (researchers, individuals).

**(i) A human rights-based approach to AI safety**

Within the ambit of the AI safety discourse in Africa, there is a need to ground the AI governance mechanism on a human rights-based approach (HRBA). This approach, a conceptual framework based on global standard on human rights[120], would help promote values such as non-discrimination, privacy, freedom of expression and the like in the design, development and deployment of AI systems. By adopting this approach to AI safety, the continent is better equipped to protect its most vulnerable members from experiencing the harmful socio-economic effects of AI.

---

[117] Ayantola Alayande, Samuel Segun, and Leah Junck. (2025 March). Emerging technology policies and democracy in Africa South Africa, Kenya, Nigeria, Ghana, and Zambia in focus. Atlantic Council. https://www.atlanticcouncil.org/wp-content/uploads/2025/03/Emerging-technology-policies-and-democracy-in-Africa.pdf

[118] Charles Asiegbu and Chinasa T. Okolo. 'How AI is impacting policy processes and outcomes in Africa'. Brookings. https://www.brookings.edu/articles/how-ai-is-impacting-policy-processes-and-outcomes-in-africa (May 2024).

[119] Sorina Teleanu, Jovan Kurbalija et al. 'Stronger digital voices from Africa: Building African digital foreign policy and diplomacy'. DiploFoundation. https://www.diplomacy.edu/wp-content/uploads/2023/01/African-digital-foreign-policy_En.pdf (November 2022).

[120] United Nations Sustainable Development Group. (n.d.) Universal Values Principle One: Human Rights-Based Approach. https://tinyurl.com/3tarsfw5

The HRBA to AI safety and security acknowledges implicitly that AI systems can undermine human rights despite the many positive contributions AI provides for the global economy. This approach calls for meaningful community participation in the development of AI., acknowledging the need for African perspectives to AI risks. Furthermore, it tasks governments and policy makers with the responsibility of developing strong legal frameworks that safeguard Africans from harm, misuse, and abuse of AI. Ultimately, it categorises AI safety not just as a technical issue but as a matter of upholding dignity, equity, and justice in digital and real-world environments.

*Outcome: AI governance across Africa is firmly rooted in human rights principles, ensuring that every AI system deployed or developed on the continent upholds and protects fundamental rights and freedoms, and does not infringe on the rights of all individuals and groups to live in safety. In practice, this would manifest as robust legal and regulatory safeguards, widespread accountability mechanisms, and meaningful inclusion of affected communities in decisions about how AI is designed, used, and governed.*

### (ii) Establishment of a dedicated centre on AI Safety in Africa

Currently, Kenya is the only African country that is part of the AI Safety Institute International Network whose members include Canada, Korea, Japan, Australia, France, Singapore, EU, UK and the US, all of which have already established a dedicated AI Safety Institute. These institutes are of varying composition – some integrated within government departments, while others are independent from government.[121] Leveraging the headway made by Kenya's involvement, we propose the establishment of a dedicated AI Safety Institute for the African continent, which tests locally developed AI models, conducts research on AI safety risks in the region, and develops policy innovations to mitigate and manage these risks.

We envisage the proposed Institute to work closely with an interdisciplinary network of scholars, practitioners, policy experts across the Continent, strengthening existing work on AI safety, and relatedly, AI security. This initiative will be designed to enhance local capabilities in addressing emerging threats, promoting justice, advancing the development of safe African AI technologies, and contributing African perspectives to global challenges and discussions. The institute should work to secure access to advanced compute, in order to conduct AI-powered evaluations of frontier AI models.

---

[121] Allen, G.C., and Adamson, G. (2024 October 30). The AI Safety Institute International Network: Next Steps and Recommendations. Center for Strategic & International Studies.
https://www.csis.org/analysis/ai-safety-institute-international-network-next-steps-and-recommendations

The Institute should aim to explore expanding understanding of large-scale AI risks in Africa and develop and test locally-designed tools for detection and mitigation. It would serve both a regional and domestic purpose, as well as an international one. Regionally and domestically, the Institute could partner with academic institutions and civil society organisations in scoping AI risks and testing solutions, set up local task teams to address particular issues, such as misinformation to manipulate public opinion, and connect with local governments and regulators to strengthen capacity to mitigate and manage AI risks. Internationally, the Institute could serve as a vehicle for ensuring African perspectives and priorities on AI safety are championed in global discussions, and contribute diverse and inclusive solutions to global challenges.

*Outcome:* *Africa has a well-resourced, trusted, and internationally recognised AI Safety Institute serving as the continental hub for cutting-edge research, policy development, testing, and capacity-building on AI safety and security. This institute would ensure that African contexts, risks, and priorities are continuously integrated into both local governance and global AI safety conversations, empowering the continent to detect, mitigate, and respond to AI-related harms with home-grown expertise and solutions.*

### (iii) Increase public literacy on AI

Africa currently has a skill and knowledge gap pertaining to AI technology,[122] making evident the need for public literacy on AI. Considering this, we propose the capacity building of students, practitioners, and policymakers, to help comprehend and address AI safety and security threats within the African context.

Several strategies for increasing public literacy on AI can be explored simultaneously to shore up the skill and knowledge gap. This can include the creation of accessible and engaging educational resources such as Massive Open Online Course (MOOC) that provide a jargon-free, linguistically diverse explanation of complex AI concepts, safety risks, and security vulnerabilities. Another approach is to integrate AI safety and security into existing curricula. This includes introducing age-appropriate AI concepts and risks into science and technology education for K-12, higher education and vocational training. Civil society organizations can also be tasked to undertake targeted AI literacy campaigns for specific sectors and populations.

---

[122] Araba Sey & Oarabile Mudongo. (2021 July 9). Case studies on AI skills capacity-building and AI in workforce development in Africa. Research ICT Africa. https://researchictafrica.net/research/case-studies-on-ai-skills-capacity-building-and-ai-in-workforce-development-in-africa/

AI-generated misinformation contents used to manipulate public opinions pose severe risks for Africa. Combating this requires creative solutions such as creating awareness campaigns and media engagements that foster a culture of critical thinking and media literacy. This strategy would require equipping the public with the skills to critically evaluate information that might be AI-generated that they encounter in the media and online.

*Outcome: People across Africa, from school children to policymakers, have the knowledge, critical thinking skills, and practical tools to understand AI, recognise its risks, and make informed choices about its use. A well-informed public and workforce would be resilient against AI-generated misinformation and better prepared to shape, adopt, and monitor safe AI technologies that align with local needs and values.*

### (iv) Develop early warning systems for Africa with inclusive benchmarks for 25+ African languages

Developing and deploying continent-wide early warning and monitoring systems for AI-driven large-scale threats is expedient for Africa. The goal of such solutions should be targeted at two interconnected challenges identified above: AI-generated fake content, and AI-enabled scaled manipulation of public opinion.

A possible approach to building such a system will combine cutting-edge AI techniques with on-ground knowledge to detect malicious content (e.g. deepfakes, coordinated disinformation campaigns) in real time across multiple channels (social media, messaging, radio). Furthermore, the development of inclusive benchmark suites for 25+ African languages is important given the language diversity on the continent. This system should be designed to prioritize linguistic diversity by incorporating local languages and dialects into data collection, risk assessment, and communication channels, ensuring no community is left behind. Inclusive benchmarks would guide the evaluation of system accuracy, fairness, and effectiveness across different linguistic and cultural contexts.

Early warning detection systems for AI risks could use multilingual models to flag incendiary narratives spreading in under-monitored languages (like Kiswahili or Oromo), and computer vision to spot deepfake images or videos before they go viral. Technically, Africa-focused early warning detection systems will advance the state-of-the-art in multilingual AI safety, developing methods to monitor and analyze content in low-resource languages that have been largely ignored by existing AI safety research. It will produce open datasets and AI models for African languages. Such a system will strengthen information integrity and resilience in several African countries, directly reducing the risk of violence or instability triggered by AI-augmented misinformation.

Civil society groups (journalists, community peacebuilders) will also benefit from access to early warnings in Africa, empowering them to respond faster to emerging issues. In the long run, wider African society benefits through more stable, truth-centered information ecosystems – an essential component of safe and democratic adoption of AI.

*Outcome: Africa has a robust, multilingual early warning and monitoring system capable of detecting AI-driven threats, from deepfakes to disinformation, in real time and across diverse languages and platforms. Communities, journalists, and authorities would be able to act swiftly to prevent the spread of harmful content, thereby protecting social cohesion, democratic processes, and public trust in information ecosystems.*

### (v) Annual AU level forum on safety and security of AI in Africa

An annual African Union–level Forum on AI Safety and Security, engaging all 55 member states, would be pivotal for coordinating, aligning and sustaining country-level and regional efforts. This recommendation aligns with the AU's Continental Strategy on AI, which calls for "an Annual Conference on AI Safety and Security in Africa" and the establishment of an expert group to assess AI's implications for peace and security.[123] The Forum should be designed as a permanent, standing mechanism under the leadership of the African Union Commission, with an initial mandate of at least ten years to ensure continuity, policy coherence, and institutional memory as the AI landscape evolves.

The Forum's core functions would include: (i) serving as a high-level platform for policymakers, regulators, researchers, security agencies, and industry leaders to exchange knowledge and harmonise standards; (ii) producing an annual State of AI Safety and Security in Africa Report, tracking emerging risks, incidents and governance gaps; (iii) developing shared policy frameworks and model regulations to guide national implementation; (iv) negotiating collectively with global technology providers for fair terms and enhanced security safeguards for African users and systems; (v) liaising with international partners such as Interpol, UN bodies, and other regional security organisations to manage transboundary AI threats, cybercrime, and misuse; and (vi) convening multi-stakeholder dialogues to build public trust, promote responsible innovation, and address socio-technical impacts specific to African contexts. The Forum should work closely with existing civil society organisations and academic groups across the Continent who are working on issues of AI risks.

---

[123] Notably, the African Union Chairperson made a statement regarding AI and its impacts on peace and security in Africa, including reference to the establishment of an Expert Group to provide advice to the AU and its member states accordingly, in March 2025.
https://au.int/en/speeches/20250320/statement-he-mahmoud-ali-youssouf-chairperson-commission-artificial-intelligence.

By institutionalising this Forum as a dedicated annual fixture, the African Union can ensure that AI safety and security remain high on national and regional agendas, strengthening Africa's collective resilience and influence in shaping the global governance of AI risks.

*Outcome: Africa has an authoritative, well-resourced, and enduring annual Forum that drives collective action, knowledge-sharing, and policy alignment on AI safety and security across all AU member states. This Forum would build a sustained pan-African community of practice, produce actionable intelligence and model policies, and amplify Africa's voice and bargaining power in global AI safety governance for at least a decade and beyond.*

**Conclusion**

Africa stands at a pivotal moment in shaping the global discourse on AI safety, with unique contextual risks and underexplored opportunities demanding urgent, locally grounded responses. This white paper has highlighted how AI risks intersect with the continent's socio-economic structures, governance challenges, and historical legacies of external dependency. While AI promises transformative benefits for African development, its safe deployment requires deliberate, coordinated action that centres human rights, cultural context, and the sovereignty of African peoples. By advancing a five-pillar action plan, encompassing rights-based policy, a dedicated AI Safety Institute, public literacy, multilingual early warning systems, and a standing AU-level forum, Africa can meaningfully contribute to and shape the emerging global AI safety architecture. This collective agenda will not only safeguard communities and ecosystems but also assert Africa's agency in steering the future of AI toward equity, resilience, and shared prosperity.